\documentclass[12pt,a4paper]{article}

\usepackage{amsmath}
\usepackage{graphicx}

\begin{document}

\title{Relativistic Point Particles and Classical Elastic Strings}
\author{C\u alin Galeriu}

\maketitle

\section*{Abstract}

{\it We extend the previous work of Olivier Costa de Beauregard regarding the
isomorphism between the equation describing the motion of relativistic point particles and 
the equation describing the static equilibrium of classical elastic strings,
by comparing the Lagrangians of these two systems.}

\section{Introduction}

The equation of motion describing the change in the four-momentum $\mathbf{P}$ of a relativistic point particle
under the action of a four-force $\mathbf{F}$ is:
\begin{equation}
d \mathbf{P} = \mathbf{F}\ d\tau,
\label{eq:costa1}
\end{equation} 
where $d\tau$ is the infinitesimal change in proper time, 
and $ds = i\ c\ d\tau$ is the length of an infinitesimal segment along the worldline of the particle.
The equation describing the static equilibrium of a classical elastic string is:
\begin{equation}
d\mathbf{T} = \mathbf{f}\ dl,
\label{eq:costa2}
\end{equation}
where $\mathbf{T}$ is the force of tension in the string, $dl$ is the length of an an infinitesimal string segment,
and $\mathbf{f}$ is the linear force density acting on that string element. 

Costa de Beauregard \cite{costa1, costa2, costa3} has noticed the isomorphism between the equations (\ref{eq:costa1}) and (\ref{eq:costa2}),
and the perfect correspondence between the physical quantities describing the three dimensional elastic string at rest
and the four dimensional relativistic particle in motion. When the string is perfectly flexible the force of tension is tangent to the string,
corresponding to a momentum-energy four-vector tangent to the worldline. A tension of constant magnitude corresponds to a particle
of constant rest mass. Other situations could also be considered, like, for example, a relativistic particle with spin angular momentum.

\section{Material Points and Infinitesimal Segments}

There is, however, one big difference between the equations (\ref{eq:costa1}) and (\ref{eq:costa2}). While the four-force acts on a 
material point particle, the linear force density acts on an infinitesimal length segment. Could we replace the relativistic point particle
with an infinitesimal element, or with an extremely small segment of length $s_o$, along the worldline of the particle? 

First we write the relativistic equation of motion (\ref{eq:costa1}) as:
\begin{equation}
\mathbf{P}_B - \mathbf{P}_A = \mathbf{F} d\tau,
\label{eq:worldline}
\end{equation}
where $A$ and $B$ are two infinitesimally close points on the worldline of the particle, with event $A$ happening before event $B$. 
The worldline is a timelike curve,
and the momentum-energy four-vectors point in the direction of the time axis.

Next we write the static equilibrium condition (\ref{eq:costa2}) as: 
\begin{equation}
\mathbf{T}_B + (- \mathbf{T}_A) + \mathbf{f} dl = 0, 
\label{eq:string}
\end{equation}
where $A$ and $B$ are two infinitesimally close points on the elastic string.
By convention, the tension vectors $\mathbf{T}_A$ and $\mathbf{T}_B$ point in the direction from $A$ to $B$.
The minus sign indicates that the tension forces
acting on the infinitesimal segment of length $dl$, at the end points, have different directions. 

The equations (\ref{eq:worldline}) and (\ref{eq:string}) are mathematically equivalent provided that 
$dl = i\ c\ d\tau$, $\mathbf{T} = i\ c\ \mathbf{P} / s_o$, and $\mathbf{f} = - \mathbf{F} / s_o$. 
Here $s_o = i\ c\ \tau_o$ is a very small constant quantity that we call the length of a particle, 
measured along its worldline. 
We also notice that the four-force acting on the material point particle and the four-force density acting on the
infinitesimal worldline segment have opposite directions. We therefore have to consistently swap attraction and repulsion between forces 
and force densities when going from one model (the point particle) to the other (the elastic string). This change of sign will
also affect the potential energy that generates the conservative forces.

From now on we will imagine the worldline of a particle as an elastic string in Minkowski space, 
under a constant tension and in static equilibrium.

\section{The Relativistic Free Particle Lagrangian}

While the non-relativistic Lagrangian of a point particle is equal to the kinetic energy minus the potential energy, 
the relativistic Lagrangian has the term
\begin{equation}
- m\ c^2 \sqrt{1 - \frac{v^2}{c^2}}
\label{eq:relativistic}
\end{equation}
in the position of the kinetic energy. This puzzling fact has been noticed by Richard Feynman, 
who writes that the term (\ref{eq:relativistic}) "is not what we have called the kinetic energy" \cite{feynman}.
We can expand (\ref{eq:relativistic}) in a power series in $v^2 / c^2$
\begin{equation}
- m\ c^2 \sqrt{1 - \frac{v^2}{c^2}} \approx - m\ c^2 + \frac{m\ v^2}{2} + ...
\label{eq:series}
\end{equation}
The constant contribution to the Lagrangian of the rest energy can be ignored, and then, 
in the non-relativistic limit $v \ll c$, we recover
the expression of the classical kinetic energy.

We distance ourselves from this interpretation, and instead declare that the term (\ref{eq:relativistic}) represents a potential energy.
The contribution of this term to the action is:
\begin{equation}
\int - m\ c^2 \sqrt{1 - \frac{v^2}{c^2}} dt = - m\ c^2 \int d\tau = i\ c\ m \int ds,
\label{eq:action}
\end{equation}
a quantity proportional to the total length of the worldline.

In a vibrating elastic string under constant tension the potential energy of an infinitesimal length element $dl$ is given by the product of the 
tension $T_o$ and the increase in length of the infinitesimal segment, relative to the equilibrium position of length $dx$. \cite{zwiebach}
\begin{equation}
T_o ( dl - dx) = T_o ( \sqrt{dx^2 + dy^2} - dx ) = T_o ( \sqrt{ 1 + ( \frac{\partial y}{\partial x} ) ^2 } - 1 ) dx
\label{PEstring}
\end{equation}
The total lenght of the string in static equilibrium, being constant, can be left inside the integral of the potential energy unsubtracted.
In order to minimize the total potential energy $\int T_o\ dl$ the elastic string under constant tension, 
in static equilibrium and not subject to any other external forces, will adopt a straight line shape between its fixed end points.
In the same way the worldline of a free particle, seen as an elastic string, will adopt a straight line shape between its fixed Minkowski 
space end points.

Richard Talman mentions that in String Theory "The strings under discussion are 
{\it ideal} (and unlike any spring encountered in the laboratory) in that their tension is independent
of their length. This being the case, the work done on a string as it is being stretched
is proportional to the change in the string's length. If the entire string rest energy is
ascribed to this work, then the string of zero length has zero rest energy and the energy of a 
string of length $l$ is equal to $T_o\ l$. Ascribing this energy to the rest mass 
of string of mass density $\mu_o$, we have $T_o = \mu_o\ c^2$. In other words, $T_o$ and $\mu_o$
are essentially equivalent parameters." \cite{talman}

We apply a similar reasoning to the worldline of a particle, seen as an elastic string.
We know that $\mathbf{T} = i\ c\ \mathbf{P} / s_o = \mathbf{P} / \tau_o$.
The magnitude of the four-momentum is $P = i\ c\ m$, where $m$ is the rest mass of the point particle. 
The magnitude of the tension is $T = i\ c\ m / \tau_o$, and the expression (\ref{eq:action}) becomes equal to $\tau_o\ T \int ds$.
We can understand in this way the formula (\ref{eq:relativistic}) of the relativistic Lagrangian of a free particle, 
which in this new interpretation relates to the potential energy of an elastic string.

\section{Directions for Further Developments}

The linear mass density of the worldline string is $\mu = m / s_o = m / (i\ c\ \tau_o)$, and the magnitude of the tension is $T = - c^2\ \mu$.
If the string were allowed to move
in hypertime (a fifth dimension), then the speed of a traveling transverse wave would be 
\begin{equation}
\sqrt{\frac{T}{\mu}} = i\ c,
\label{eq:speed}
\end{equation}
which, coincidentally or not, is the speed (the magnitude of the four-velocity) of the human conscience 
moving along the worldline of the human being. 

The main goal of this article was to show that the variational method commonly used with classical elastic strings 
is also applicable to worldline strings.
Here we have discussed only the simplest case, but the variational method can also be applied to more complex examples.
We could have situations when the tension in the woldline string 
is not constant (the rest mass of the particle is not constant), or when the worldline string is subject to external force
densities (the particle is subject to a potential field that generates conservative forces).

While the generally accepted view is that the magnitude of the
four-momentum of a particle is constant (the rest mass is constant), and as a result the magnitude of the tension in the 
worldline string is constant, this is
not an absolutely necessary requirement. It was already noticed by Hugo Tetrode \cite{tetrode} a long time ago that in the most general situation 
the rest masses of the particles may depend upon their four-accelerations. When the rest mass is not constant
we have to write $\int m\ ds$ instead of $m \int ds$ in the expression of the relativistic action. This is indeed what we expect from 
the variational treatment of an elastic string under a variable tension.

\section*{Acknowledgments}

The author is indebted to David H. Delphenich for 
translating from German into English the article by Hugo Tetrode. \cite{tetrode}

\end{document}